\documentclass[fleqn,10pt]{wlscirep}
\title{In-gap corner states in core-shell polygonal quantum rings}

\author[1,2,3*]{Anna Sitek}
\author[4]{Mugurel \c{T}olea}
\author[4]{Marian Ni\c{t}\u{a}}
\author[5,6]{Lloren\c{c} Serra}
\author[1]{Vidar Gudmundsson}
\author[2]{Andrei Manolescu}
\affil[1]{Science Institute, University of Iceland, Reykjavik, IS-107, Iceland}
\affil[2]{School of Science and Engineering, Reykjavik University, Reykjavik, IS-101, Iceland}
\affil[3]{Department of Theoretical Physics, Faculty of Fundamental Problems of Technology, Wroclaw University of Science and Technology, Wroclaw, 50-370, Poland} 
\affil[4]{National Institute of Materials Physics, Bucharest-Magurele, P.O. Box MG-7, Romania}
\affil[5]{Institute of Interdisciplinary Physics and Complex Systems IFISC (CSIC-UIB), Palma de Mallorca, E-07122, Spain}
\affil[6]{Department of Physics, University of the Balearic Islands, Palma de Mallorca, E-07122, Spain}

\affil[*]{anna.sitek@pwr.edu.pl}

\usepackage{soul}
\usepackage{bm}

\usepackage{xcolor}

\newcommand{\blue}{\textcolor{black}}    

\begin{abstract}
We study Coulomb interacting electrons confined in polygonal quantum rings. 
We focus on the interplay of localization at the polygon corners and Coulomb 
repulsion. Remarkably, the Coulomb repulsion allows the formation of 
\textit{in-gap states}, i.e., corner-localized states of electron pairs or clusters  
shifted to energies that were forbidden for non-interacting electrons, but below 
the energies of corner-side-localized states. 
We specify conditions allowing optical excitation to those states.
\end{abstract}

\begin{document}

\flushbottom
\maketitle
%
%
\thispagestyle{empty}

\section*{Introduction}
Core-shell quantum wires are vertically grown nanoscale structures 
consisting of a core which is covered by at least one layer of a different
material (shell). 
Recently these structures attracted considerable 
attention as building blocks of quantum nanodevices 
\cite{Krogstrup13,Tang11,Kim15,Xiang06,Nguyen14,Saxena13,Ho15,Thierry12,Ibanes13,Peng15}.
A characteristic feature of core-shell systems is a non-uniform carrier 
distribution in different parts of the wire  
\cite{Jadczak14,Bertoni11,Royo13,Royo14,Royo15,Fickenscher13,Shi15}. 
It is a consequence of the polygonal cross
section which most commonly is 
hexagonal \cite{Blomers13,Rieger12,Haas13}, but may also be 
triangular \cite{Qian04,Qian05,Baird09,Heurlin15,Dong09,Yuan15}, 
square \cite{Fan06}, or dodecagonal \cite{Rieger15}.
Some of the properties of those wires, 
such as the band alignment \cite{Pistol08,Wong11}, 
may be controlled to a high extent. An appropriate combination of 
sample size and shell thickness allows 
to induce electron concentration on the shell area \cite{Blomers13}. 
Moreover, the present technology allows for etching out the core 
part and producing nanotubes \cite{Rieger12,Haas13}.

Both nanowires and nanotubes may be viewed as polygonal quantum rings if
they are sufficiently short, i.e., shorter than the electron wavelength
in the growth direction. In this geometry the single-particle states with 
the lowest energy are localized in the corners of the polygon and are separated 
by a gap from the states localized on the sides.  The gap can be of tens of 
meV or larger, depending on the shape of the polygon \cite{Sitek15,Sitek16}.

The single-particle energy levels of a polygonal quantum
ring are two- and fourfold degenerate and their arrangement is 
determined only by the number of polygon vertices.
Similarly to the case of bent parts of quantum wires \cite{Sprung,Lent,Sols,Wu,Wu2,Wu3,Vacek,Xu}, 
in the corner areas of polygonal quantum rings effective quantum wells are formed and thus 
low energy levels localize between internal and external boundaries. 
The number of such corner states is the number of vertices times two spin orientations. 
An energy gap may separate single-particle corner states 
from higher-energy states,  the latter being distributed 
over the polygon sides \cite{Ballester12,Estarellas15,Sitek15}.

In this paper we extend the single-particle model of 
Refs.\ \cite{Sitek15} and\ \cite{Sitek16} to systems of few 
Coulomb interacting electrons. We show 
how this coupling allows the formation of states corresponding to 
electron pairs, or larger clusters, that localize on the corners
and whose energies lie in the  gap between
corner and corner-side states of the uncoupled system. 
We focus on the formation and excitation of those many-body  \textit{in-gap states}, 
with particular emphasis on their fingerprints in optical absorption.  
As general motivations to study in-gap states in polygonal rings we mention their 
potential application in quantum information devices, 
exploiting the corner occupation as information unit, or their use as 
quantum simulators of discrete lattice models \cite{Georgescu14}.

\section*{Results}

Below we analyse systems of up to five electrons confined in triangular, square and 
hexagonal quantum rings. We take into account only conduction band electrons and 
neglect the \blue{valence} levels or assume that the \blue{valence band is} fully occupied. 
This situation can be achieved and controlled with a nearby metal gate, like in 
the single electron tunnelling experiments with quantum dots, or possibly with 
an STM tip.

For all of the analysed polygons the external 
radius ($R_{\mathrm{ext}}$) and side thickness ($d$) are fixed to 50 and 12 nm, 
respectively. The rings we describe are in fact short prismatic structures. We 
consider all electrons in the lowest mode in the growth direction, and the energy 
interval up to the second mode larger than the energy gap between the lateral modes 
of corner and side type. This gap varies from $27.5$ meV for triangular rings 
to $4.1$ meV for hexagonal samples. Assuming, e.g., the length (or height) in the 
growth direction equal to $d$, and the InAs parameters, an estimation of the energy 
separation between the two lowest longitudinal modes, using the simple quantum box 
model, gives 342 meV, which is above the energy range we are interested in this paper.
\blue{In other words, the prism length is not important as long as it is short enough
to guarantee sufficiently large separation between the two lowest longitudinal modes, i.e.,
larger than the energy range due to the lateral confinement.}

\subsection*{\label{sec:triangle} Many-body states for a triangular ring}


\begin{figure}[ht]
\centering
 \includegraphics[scale=0.67]{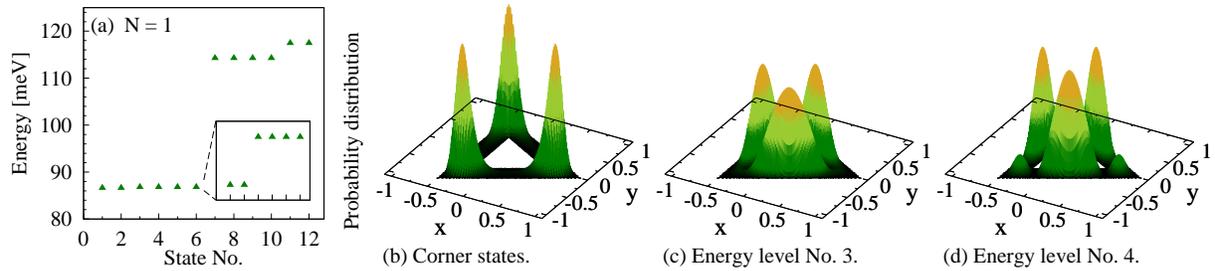} 
\caption{\blue{Single-particle quantities for a triangular ring.
         (a) The $12$ lowest states arranged into $4$ energy levels, 
         the inset shows the degeneracy of corner states.
         (b-d) Probability distributions associated with the 
         energy levels shown in Fig. (a).
         The $x$ and $y$ coordinates are in units of $R_{\mathrm{ext}}$}. }
 \label{fig:single}
\end{figure}

The ground state of a single electron confined in a symmetric triangular 
quantum ring is twofold degenerate and is followed by a sequence of alternating 
pairs of four- and twofold degenerate levels, \blue{Fig.\ \ref{fig:single}(a)}.
For the analysed 12 nm thick ring the lowest six states
are localized in the corners, with well-separated probability peaks in
the areas between internal and external boundaries, and vanishing
probability distributions on the sides of the ring, \blue{Fig.\ \ref{fig:single}(b)}.  
The higher six states are distributed mostly over the sides of the
triangle, with only little coverage of the corners, 
\blue{Figs.\ \ref{fig:single}(c) and\ \ref{fig:single}(d)}. 
In addition, corner and side states are separated by an energy gap $\Delta_t=27.5$ meV 
($t$ meaning {\it triangle}), which rapidly increases if the aspect 
ratio $d/R_{\rm{ext}}$ is reduced \cite{Sitek15,Sitek16}.

\begin{figure}[ht]
\centering
 \includegraphics[scale=0.67]{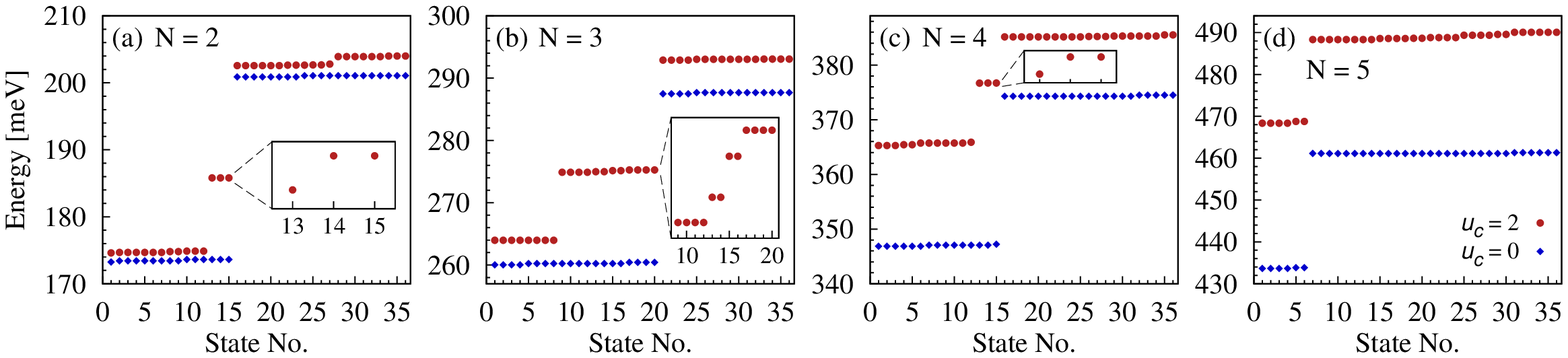} 
\caption{Energy levels for a triangular ring.
         The number of confined electrons ($N$) is shown in each figure and
         the interaction parameters shown in Fig.\ (d) are valid for all figures.
         \blue{In the insets to panels (a-c) we show the fine structure of the in-gap states.}} 
 \label{fig:energy}
\end{figure}

\begin{figure}[ht]
\centering
 \includegraphics[scale=0.67]{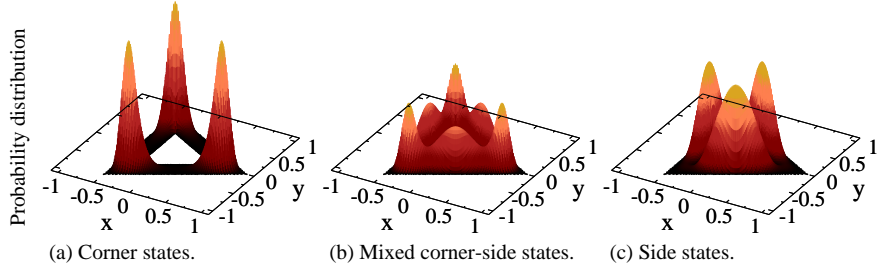} 
\caption{\blue{Two-particle lateral localization. Probability distributions for two 
         electrons in corner \blue{(including the in-gap)} 
         states  (a), in mixed corner-side states (b) and in side states (c).
         The $x$ and $y$ coordinates are in units of $R_{\mathrm{ext}}$.}}       
 \label{fig:localization}
\end{figure}

For $N=2$  non-interacting electrons  confined in our triangular ring
the low energy states form nearly dispersionless groups (flat bands),
of fifteen corner states followed by thirty-six mixed corner-side
states, represented by the blue diamonds in Fig.\ \ref{fig:energy}(a).
Clearly, the two flat bands are separated by approximately $\Delta_t$.  
The lowest group of many-body states has probability distributions 
qualitatively similar to the single-particle corner-localized states,
Fig.\ \ref{fig:localization}(a). The levels above the gap contain 
contributions from both, corner- and side-localized, single-particle 
states and thus are associated with mixed corner-side probability 
distributions, Fig.\ \ref{fig:localization}(b).
The third group of states is built up of only side-localized 
single-particle states [states 7-12 in Fig.\ \ref{fig:single}(a)] 
and associated with probability
distributions of that kind,  Fig.\ \ref{fig:localization}(c).

The Coulomb interaction between the two electrons does not change
qualitatively the charge distributions around the polygon shown in Fig.\
\ref{fig:localization}, as long as the Coulomb energy is smaller than
$\Delta_t$.  Instead, the energy spectrum differs qualitatively from
the case of non-interacting particles, as shown in Fig.\ \ref{fig:energy}(a).
The corner states 1-12 are only slightly shifted up, indicating that
they correspond to electrons situated in different corners, in singlet
or triplet spin configurations.  The ground state is singlet and
non-degenerate, the next energy levels are  triplet sixfold, singlet
twofold, and triplet threefold degenerate, respectively.  These twelve
states are only slightly spread within a narrow energy range, of
about 0.25 meV.  

Contrary to the behaviour of the first group of states, the next three corner
states, 13-15, are shifted to higher energies, within the 
former gap of forbidden energies for non-interacting
particles, Fig.\ \ref{fig:energy}(a).  These in-gap
states correspond to both electrons occupying the same corner area,
with spin singlet configuration, and with an increased Coulomb energy.
The localization of such states is still like in Fig.\
\ref{fig:localization}(a).  Obviously, the charge density is 
equally distributed between the three symmetric corners.

The energy spectrum changes with the number of electrons.  First of
all it moves up due to the increased Coulomb energy.  If $N=3$ there are
twenty many-body corner states, and twelve of them are lifted into the
gap, Fig.\ \ref{fig:energy}(b). These states correspond to situations
when two electrons of different spin occupy the same corner area while
the third electron is localized around one of the two other corners.
The energy spectrum for $N=4$, Fig.\ \ref{fig:energy}(c),
resembles the one for $N=2$.  This is a
kind of particle-hole symmetry in the Fock space associated to the six
single-particle corner states.  In both cases three states are shifted
into the gap, which for $N=4$ correspond to two corners doubly occupied
or one corner unoccupied.  When $N=5$, there are only six states
associated with purely corner-localized probability distributions, with
two corner areas occupied by a pair of electrons while the fifth electron
stays on the third corner.  As for $N=1$, no in-gap state
exists in this case, Fig.\ \ref{fig:energy}(d).

Degeneracies of the two-particle in-gap states, \blue{inset to} Fig.\ \ref{fig:energy}(a),
and the ones associated with one unoccupied corner area, \blue{inset to} Fig.\
\ref{fig:energy}(c), reproduce the degeneracy of the lowest
single-particle levels with respect to spin \blue{[inset to Fig.\ \ref{fig:single}(a)]} 
which in this case is conserved and thus the degeneracy of these levels is only of the orbital
origin. This is not the case when some of the electrons are unpaired,
such systems are spin polarized and some of their in-gap levels are
fourfold degenerate, Fig.\ \ref{fig:energy}(b).
%
In the presence of a magnetic field normal to the surface of the polygon 
the degeneracies are lifted. Still, the corner localization is not affected, 
as long as the Zeeman energy is smaller than the energy gap ($\Delta_t$), 
such that the mixing of corner and corner-side states is not significant.

\subsubsection*{\label{sec:hubbard} Comparison with the Hubbard model}

The corner localization of the low-energy states suggests that we can 
obtain some insight from a Hubbard model with on-site Coulomb energy 
$U$ and inter-site hopping energy $t$. Nevertheless, even for such a simplified model, 
only the simplest case of two electrons in triangular ring can be solved 
analytically \cite{Korkusinski07}. The solution consists of 9 triplet states 
insensitive to interaction and 6 singlet states. Two singlets are non-degenerate
while other two are both twofold degenerate, with energies

\begin{equation}
E_{1,2}^{\uparrow\downarrow}=\frac{1}{2} \left(U + 2t \pm \sqrt{U^2-4tU+36t^2}\right) , 
\end{equation}
\begin{equation}
E_{3,5}^{\uparrow\downarrow}=E_{4,6}^{\uparrow\downarrow}
=\frac{1}{2} \left(U - t \pm \sqrt{U^2+2tU+9t^2}\right) .
\end{equation}
 
One can notice that in the limit of $U \gg \vert t\vert$ the energies 
$E_{1,3,4}^{\uparrow\downarrow} \to U$,
whereas $E_{2,5,6}^{\uparrow\downarrow} \to 0$.  
At the same time all triplet states have by 
default zero energy. This spectrum is qualitatively similar to the energies 
obtained earlier for the corner states of the thin triangle \cite{Sitek16}. 
Therefore the energy difference between the in-gap states
and the ground state is approximately the $U$ parameter of the Hubbard 
model.   The fine structure, i.e., splitting of the three in-gap states,
is $E_1^{\uparrow\downarrow}<E_3^{\uparrow\downarrow}=E_4^{\uparrow\downarrow}$, 
for $t<0$, in agreement with the previous results. 

For three electrons on the triangle numerical results within the Hubbard 
model confirm twelve states shifted up by the Coulomb 
repulsion. However, for $U\gg \vert t\vert$, their fine structure consists of three degenerate 
levels (4+4+4) instead of four (4+2+2+4) shown in Fig.\ \ref{fig:energy}(b).

A simple connection between the main model and the Hubbard 
one can be made if we keep in mind that,
for a triangle with side thickness $d$, there
is a simple relation between the energy of the in-gap states and the
pair Coulomb energy $u_c$.  The in-gap states have an extra energy
$E_c=u_c/\lambda$, where $\lambda$ is the distance between the two
electrons sitting on the same corner, in units of $R_{\rm{ext}}$.  Since the
wave functions must vanish at the lateral boundaries $\lambda$ should
be smaller than $d$.  From the energy data we estimate $\lambda=0.6d$.
Clearly $E_c$ is linear with $u_c$.   From the results of the Hubbard
model, if we assume no hopping between corners ($t=0$),
then we have $U=E_c$.

\subsection*{\label{sec:square_hex} Many-body states for square and hexagonal rings}

\begin{figure}[ht]
\centering
 \includegraphics[scale=0.67]{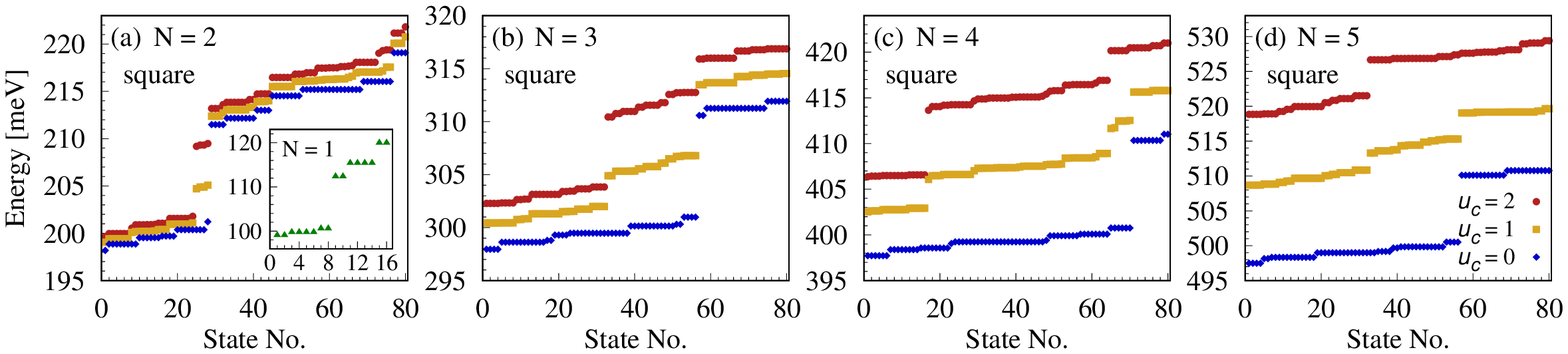} 
\caption{Energy levels for a square ring. 
         The number of confined electrons ($N$) is shown in each figure and
         the interaction parameters given in Fig. d are valid for all figures.}
 \label{fig:energy_S}
\end{figure}


The single-particle energy gap between corner and side states decreases
with increasing number of corners \cite{Sitek15}.  For a $12$ nm thick
square ring this gap splitting is still sizeable, $\Delta_s=11.8$ meV
($s$ meaning {\it square}), as shown in Fig.\ \ref{fig:energy_S},
but smaller than in the case of 12 nm thick triangle.  As a result, the
range of Coulomb interaction strengths $u_c$ allowing formation of  in-gap
states reduces.  For example, in Fig.\ \ref{fig:energy_S} we show results
for $u_c=1$ and $u_c=2$.
For $N=2$ four in-gap states are created, again corresponding to spin-singlet
pairs occupying the same corner.  The degeneracy of these states is 1,2,1 (in
energy order). As for the triangle occupied by 2 or 4 electrons, they reproduce 
the degeneracy of the single-particle corner states up to a spin factor of two.
The spectra become more complex with increasing the number of electrons.
Still, interesting effects occur, for example for $N=4$.  In this case
two groups of in-gap states may be formed.  One group with one singlet
pair at one corner and the other two electrons in different corners, and
another group, with a higher energy, but still in the gap, with two pairs
at two corners. This situation is shown in Fig.\ \ref{fig:energy_S}(c)
for $u_c=1$. In the case of $N=5$ only one group of in-gap states is 
formed, Fig.\ \ref{fig:energy_S}(d).
When the interaction strength is increased to $u_c=2$ then all 
states corresponding to two corner areas occupied by a singlet pair 
are shifted to energies above the gap and mix up with levels 
associated with corner-side-localized probability distributions
[red circles in Figs.\ \ref{fig:energy_S}(c) and \ \ref{fig:energy_S}(d)].


\begin{figure}[ht]
\centering
 \includegraphics[scale=0.67]{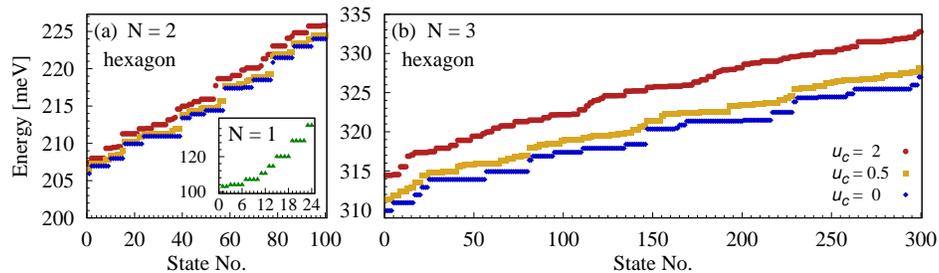} 
\caption{Energy levels for a hexagonal ring.
         The number of confined electrons ($N$) is shown in each figure and
         the interaction parameters given in Fig. b are valid for both figures.}
 \label{fig:energy_H}
\end{figure}

For a hexagonal ring of $12$ nm thickness the energy interval between the 
single-particle corner and side states is $\Delta_h=4.1$ meV, which is  
comparable to the energy spacing within these two groups of states.
\blue{The reason is that the localization of the electrons is much 
weaker for the corner angle of 120 deg than for the previous cases of 
60 and 90 deg.
This results in the overlap of the energy domains of purely corner 
and mixed corner-side many-body 
states even for non-interacting particles. Consequently there is no energy 
gap above the many-body corner states in the examples shown in Fig.\ \ref{fig:energy_H}. 
The Coulomb interaction does 
not affect considerably the energy structure of such samples, it only shifts
the levels to higher energies and changes the order of some states. 
However, the single-particle energy gap $\Delta_h$ increases with decreasing 
aspect ratio \cite{Sitek16}, and thus, for sufficiently thin
rings many-body corner states could be energetically separated from the other states as 
in the case of triangular and square samples. For such rings the} 
Coulomb interaction either mixes higher corner and corner-side states
or, if it is sufficiently weak, it reorganizes the corner states into 
groups corresponding to each number of close-by singlet pairs 
\blue{and different spatial separations of the particles}. 
In \blue{all those} cases in-gap levels cannot be identified as before. 
Still, theoretically, with a \blue{very} low aspect ratio and a weak 
interaction such states could be obtained.

\section*{\label{sec:absorption} Electromagnetic absorption}

\begin{figure}[ht]
\centering
 \includegraphics[scale=0.67]{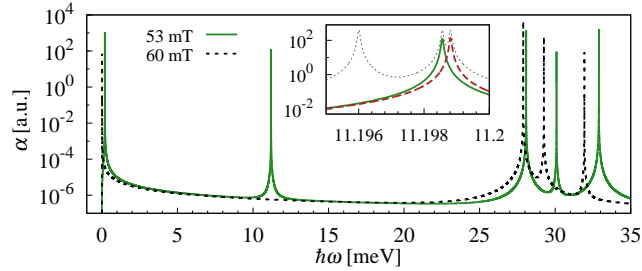}
\caption{Absorption spectrum.
         Absorption coefficients associated with the excitation of the 
         ground state of a pair of interacting electrons confined on a 
         triangle impinged with clockwise polarized 
         electromagnetic field and exposed to a magnetic field 
         of 53 and 60 mT, close to the singlet-triplet ground state transition.
         Inset: Density of states (grey dotted) and absorption coefficients
         associated with clockwise (green solid) and counter-clockwise (red dashed)
         polarized electromagnetic field for the in-gap states. }
\label{fig:absorption}
\end{figure}

A single electron confined in a triangular quantum ring simultaneously
exposed to a static magnetic field and circularly polarized
electromagnetic field may be excited from its ground state to only four
higher states within the 12 lowest states.
Two of these states are associated
with corner-localized probability distributions and originate from 
splitting of the second level of the degenerate system ($B=0$) while the
other two states belong to the group of side-localized states and merge
into the third (fourfold degenerate) level when the magnetic field is
removed. Two transitions, one to a state below and the other one to a
state above the gap separating corner- from side-localized states, take
place in the presence of each polarization type \cite{Sitek15,Sitek16}.

Considering now a pair of Coulomb interacting electrons in the ground
state (and $u_c=2$), it may be excited with clockwise polarized
electromagnetic field to one of the corner states with nearby energy,
to one of the in-gap corner states, or to three states associated with
mixed corner-side probability distributions (green solid lines in Fig.\
\ref{fig:absorption}).  One of the reasons why so many transitions are
forbidden is that we do not take into account spin-orbit interaction and thus
restrict transitions to pairs of states associated with the same spin.
Moreover, some single-particle transitions are blocked due to wave
function symmetry \cite{Sitek15,Sitek16}. Consequently, the few allowed
many-body transitions are the ones for which the corresponding dipole
moment matrix elements contain the appropriate combinations of pairs of
optically accessible single-particle states.  Similar transitions to those
shown in Fig.\ \ref{fig:absorption}, but to different states, take place when
the sample is excited with counter-clockwise polarized electromagnetic
field.

\begin{figure}[ht]
\centering
 \includegraphics[scale=0.67]{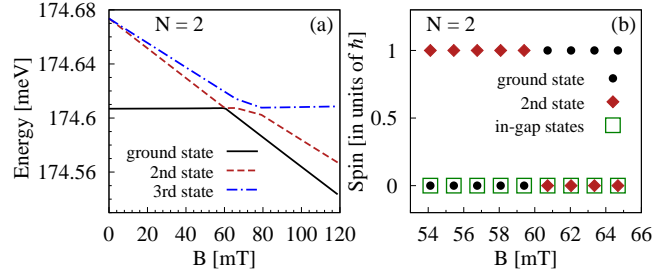}
\caption{Effect of a magnetic field. (a) The three lowest energy states and
         (b) spin associated with the ground state, the first excited 
         state, and the in-gap states of two electrons 
         confined in a triangular ring versus magnetic field.}
 \label{fig:magnetic_spin}
\end{figure}

In particular, in the inset to Fig.\ \ref{fig:absorption} we show the
part of many-body absorption spectrum and density of states associated
with the in-gap states for $B=53$ mT.  Two out of these states may be
optically reached from the ground state, each one when the sample is
impinged with differently polarized electromagnetic field (green solid
and red dashed lines).  The two final in-gap states are spin singlets,
but split by the orbital effect of the magnetic field, and thus the
transitions induced with clockwise and counter-clockwise polarization
merge when the magnetic field is removed ($B=0$).


These in-gap states may be optically excited from the ground state only
for a sufficiently low magnetic field, as long as the ground state is
spin singlet, as it is for $B=0$. In our case this regime corresponds
to $B<60$ mT.  This situation changes when the external field reaches
$60$ mT.  At this point a spin polarized state becomes the ground
state, originating in a spin triplet at $B=0$, as shown in Fig.\
\ref{fig:magnetic_spin}(a).  As seen in Fig.\ \ref{fig:magnetic_spin}(b)
the two lowest states are associated with different spin, so when the
levels cross the spin of the ground state changes, while the in-gap
states remain spin singlets over the transition range [green squares in
Fig.\ \ref{fig:magnetic_spin}(b)].  Consequently, the matrix elements of
the dipole moment between the \textit{new} ground state and the in-gap
states vanish, together with the optical coupling, as indicated
by the black dashed line in Fig.\ \ref{fig:absorption}.  The
in-gap states may still be optically excited in the presence of higher
magnetic fields, but from different initial states. Such states
evolve from the ground state at $B=0$, e.g. the second state for $60<B<80$ T, 
or the third state for $B>80$ mT [red dashed or blue dash-dotted lines in 
Fig.\ \ref{fig:magnetic_spin}(b), respectively], or possibly
from other spin singlet states.

\section*{Discussion}

We studied energy levels, localization, and optical absorption 
of systems of few electrons confined in polygonal quantum rings. 
If the numbers of corners and particles allow formation of purely 
corner-localized states corresponding to close-by
singlet pairs, then the states associated with particular number 
of such pairs form separated groups which are shifted to higher 
energies and either form in-gap states in the energy ranges forbidden
to non-interacting electrons, or mix up with levels associated with
corner-side-localized probability distributions.  An applied magnetic
field (in the range of tens of mT) may lead to ground state spin change,
for instance for two electrons towards spin alignment, while the in-gap
states remain singlets, which results in blocking of the excitations
towards the in-gap states. This provides a possibility of experimental
testing of the sample shape and interaction strength with a contactless
control of the absorption process. The presence of spin-orbit interaction
in the core-shell structure, which would remove the spin selection rules,
can also be experimentally tested.  In general such a structure
has a non-uniform dielectric constant, and thus the effective Coulomb
potential is different from the standard $1/r$ form used in our model.
Still, as long as the pairwise Coulomb energy is smaller than the gap between
corner and side states our results remain qualitatively valid.

\section*{Methods}

\begin{figure}[ht]
\centering
 \includegraphics[scale=0.67]{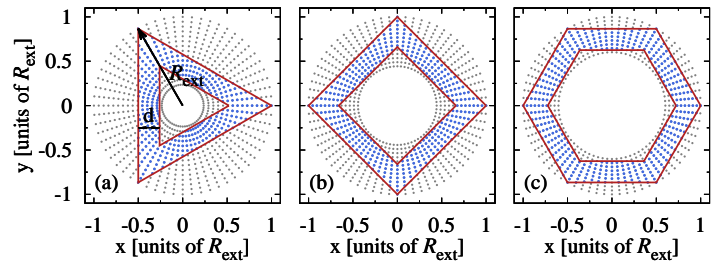}
\caption{Sample models: Different polygonal constraints (red solid lines) 
         applied on a polar grid (grey points) which is further reduced 
         to sites situated only between the boundaries (blue points). 
         The black arrows indicate the external radius of the polar grid 
         and thus of the polygonal rings ($R_{\mathrm{ext}}$) and side 
         thickness ($d$).        
         For visibility we reduced the number of site points.}
\label{fig:samples_TSH}
\end{figure}

Our model of a polygonal quantum ring is based on
a discrete polar grid \cite{Daday11} 
on which we superimpose polygonal constraints and restrict to sites situated 
between the boundaries (Fig.\ \ref{fig:samples_TSH}).
The single-particle Hamiltonian is 
\begin{eqnarray*}
\label{Hamiltonian_single_1}
 H = \frac{\left(-i\hbar\nabla+e\bm{A}\right)^{2}}{2m_{\mathrm{eff}}}
-g_{\mathrm{eff}}\mu_{\mathrm{B}}\,\sigma_{z}\,B\; ,
\end{eqnarray*}
where $\bm{A}$ is the vector potential of an external 
magnetic field $B$ normal to the ring plane ($x,y$), $m_{\mathrm{eff}}$ the effective mass 
of the ring material, $g_{\mathrm{eff}}$ the effective g-factor and $\sigma_{z}$ 
the $z$th Pauli matrix.  The Hilbert space associated with our
polar lattice is spanned by the vectors $\vert k j \sigma\rangle$, including
the radial ($k$) and angular ($j$) coordinates, and the spin ($\sigma$).
The matrix elements $\langle kj\sigma\vert H\vert
k'j'\sigma'\rangle$ are used to obtain single-particle eigenvalues
$E_{a}$ and eigenvectors $\psi_{a}$ by numerical diagonalization
\cite{Daday11,Sitek15}.

The many-body Hamiltonian of interacting electrons is 
\begin{equation*}
\label{Hamiltonian_many}
\hat {H}=\sum\limits_{a}E_{a}a^{\dagger}_{a}a_{a}+\frac{1}{2}\sum\limits_{abcd}V_{abcd}a^{\dagger}_{a}a^{\dagger}_{b}a_{d}a_{c}\; ,
\end{equation*}
where operators $a^{\dagger}_{a}$ and $a_{a}$ create and annihilate, respectively, 
an electron in the single-particle eigenstates, while $V_{abcd}$ are 
the Coulomb integrals,
\begin{equation*}
\label{Coulomb_int}
V_{abcd} = \langle \psi_{a}\psi_{b}\vert \frac{e^{2}}{\kappa\vert \bm{r} - \bm{r}' \vert} \vert\psi_{c}\psi_{d} \rangle,
\end{equation*}
where $\kappa$ is the material dielectric constant and $|\bm{r}-\bm{r}'|$
the spatial separation of an electron pair. The many-body sates are
obtained by (exact) diagonalization of $\hat{H}$ in a truncated Fock
space, typically including a number of single-particle states  up to four
times the number of polygon corners, and several thousands of grid points.

We calculate the absorption coefficients for the many-body system 
in the dipole and low
temperature approximations. As derived e.g.  
in Ref.\ \cite{Chuang95}, they are
\begin{equation*}
\label{absorption_coeff}
 \alpha(\hbar\omega) = {\cal{A}}\,\hbar\omega\,\Gamma\sum_{f}
 \frac{|\,\langle f|\bm{\varepsilon}\cdot\bm{p}|i\rangle\,|^2}
 {\left[\,\hbar\omega
-\left({\cal{E}}_{\mathrm{f}}-{\cal{E}}_{\mathrm{i}}\right)\,\right]^2
+ \left(\Gamma/2\right)^2}\ ,
\end{equation*}
where ${\cal{A}}$ is a constant amplitude, $\bm{\varepsilon}=\left(1,\pm
i\right)/\sqrt{2}$ correspond to circular polarizations of the
electromagnetic field, $\bm{p}$ is the electric dipole moment, $\Gamma$
a phenomenological broadening, and ${\cal{E}}_{\mathrm{i,f}}$ are
the energies corresponding to the initial ($|i\rangle$) and final
($|f\rangle$) many-body states.


Our results were obtained for triangular, square and hexagonal quantum rings. 
For all of the analysed polygons the external radius ($R_{\mathrm{ext}}$)
and side thickness ($d$) are fixed to 50 and 12 nm, respectively.
In the numerical calculations we use as energy unit 
$t_0\equiv\hbar^2/2m_{\rm eff}R_{\rm {ext}}^2$.  
Typically, we consider InAs as reference material, with
$m_{\mathrm {eff}}=0.023$, $g_{\mathrm {eff}}=-14.9$, and $\kappa=15$. 
The strength of the Coulomb interaction of an electron pair is defined by 
the parameter $u_c=(e^2/\kappa R_{\rm{ext}}) / t_0 $, which is about 2.9 for InAs.  
In our calculations we consider $u_c=1$, $0.5$ and $2$.
In order to resolve the fine structure of the absorption spectra 
we use $\Gamma = 0.066$ meV.


\section*{Acknowledgements}
This work was financed by the Research Fund of the University of 
Iceland and the Icelandic Research Fund.

\end{document}